\documentclass[conference]{IEEEtran}
\IEEEoverridecommandlockouts

\usepackage{amsmath,amssymb,amsfonts}   %
\usepackage{textcomp}                   %

\usepackage{graphicx}                   %
\usepackage{xcolor}                     %

\usepackage{array}                      %
\newcolumntype{H}{>{\setbox0=\hbox\bgroup}c<{\egroup}@{}} %
\usepackage{booktabs}                   %
\usepackage{siunitx}                    %
  \DeclareSIUnit{\dBm}{dBm}
  \DeclareSIUnit{\dBi}{dBi}
  \DeclareSIUnit{\dBsm}{dBsm}

\usepackage{soul}                       %

\usepackage[backend=biber,
            language=english,
            autolang=other,
            style=ieee,
            maxcitenames=1,mincitenames=1
            ]{biblatex}       %
\AtBeginBibliography{\footnotesize}

\usepackage[dvipsnames,svgnames,table]{xcolor} %
\usepackage[hidelinks]{hyperref}               %
\usepackage[capitalize]{cleveref}                          %

\usepackage{algorithmic}                %

\usepackage{pgfplots}                   %
  \pgfplotsset{compat=newest}
\usepgfplotslibrary{groupplots,dateplot} %
\usepackage{pgfplotstable}              %
  \pgfplotsset{compat=1.18}              %

\usepackage{circuitikz}
\usetikzlibrary{%
  patterns,          %
  shapes.arrows,     %
  spy,               %
  external,          %
  calc,              %
  shadings,          %
  shadows.blur,      %
  decorations.pathreplacing %
}

\usepackage[xindy,nonumberlist]{glossaries} %
  \makenoidxglossaries            %
  
\newacronym{2d}{2D}{two-dimensional}
\newacronym{3d}{3D}{three-dimensional}
\newacronym{3gpp}{3GPP}{3rd generation partnership project}
\newacronym{3msk}{3MSK}{3-level minimum-shift-keying}
\newacronym{4g}{4G}{fourth-generation}
\newacronym{5g}{5G}{fifth-generation}
\newacronym{6dof}{6DoF}{six degrees of freedom}
\newacronym{6g}{6G}{sixth-generation}
\newacronym{abp}{ABP}{authentication by personalisation}
\newacronym{ac}{AC}{alternating current}
\newacronym{aci}{ACI}{adjacent channel interference}
\newacronym{aclr}{ACLR}{adjacent channel leakage ratio}
\newacronym{acpr}{ACPR}{adjacent channel power ratio}
\newacronym{adc}{ADC}{analog-to-digital converter}
\newacronym{adr}{ADR}{adaptive data rate}
\newacronym{aec}{AEC}{aluminum electrolytic capacitor}
\newacronym{aes}{AES}{Advanced Encryption Standard}
\newacronym{afe}{AFE}{analog front end}
\newacronym{ag}{AG}{array gain}
\newacronym{agc}{AGC}{automatic gain controller}
\newacronym{agps}{A-GPS}{Assisted Global Positioning System} %
\newacronym{ai}{AI}{artificial intelligence}
\newacronym{alk}{ALK}{Alkaline}
\newacronym{am}{AM}{amplitude modulation}
\newacronym{amam}{AM/AM}{amplitude modulation to amplitude modulation}
\newacronym{amc}{AMC}{automatic modulation classification}
\newacronym{amp}{AMP}{approximate message passing}
\newacronym{ampm}{AM/PM}{amplitude modulation to phase modulation}
\newacronym{aoa}{AOA}{angle-of-arrival}
\newacronym{aod}{AOD}{angle-of-departure}
\newacronym{ap}{AP}{access point}
\newacronym{api}{API}{application program interface}
\newacronym{apt}{APT}{acoustic power transfer}
\newacronym{apu}{APU}{access point unit}
\newacronym{ar}{AR}{augmented reality}
\newacronym{arima}{ARIMA}{Auto-Regressive Integrated Moving Average}
\newacronym{arp}{ARP}{Antenna Reference Point}
\newacronym{asic}{ASIC}{application specific integrated circuit}
\newacronym{ask}{ASK}{amplitude-shift keying}
\newacronym{at}{AT}{ATtention}
\newacronym{auv}{AUV}{autonomous underwater vehicle}
\newacronym{awg}{AWG}{arbitrary waveform generator}
\newacronym{awgn}{AWGN}{additive white Gaussian noise}
\newacronym{baw}{BAW}{bulk acoustic wave}
\newacronym{bb}{BB}{base-band}
\newacronym{bc}{BC}{backscatter communication}
\newacronym{bcjr}{BCJR}{Bahl-Cocke-Jelinek-Raviv}
\newacronym{bd}{BD}{backscatter device}
\newacronym{be}{BE}{Belgium}
\newacronym{ber}{BER}{bit error rate}
\newacronym{bf}{BF}{beamforming}
\newacronym{bga}{BGA}{ball grid array}
\newacronym{bldc}{BLDC}{brushless DC}
\newacronym{ble}{BLE}{Bluetooth Low Energy}
\newacronym{bler}{BLER}{block error rate}
\newacronym{bod}{BOD}{Brown-Out Detection}
\newacronym{bom}{BOM}{bill of materials}
\newacronym{bpsk}{BPSK}{binary phase-shift keying}
\newacronym{brp}{BRP}{beam refinement process}
\newacronym{bs}{BS}{base station}
\newacronym{bu}{BU}{booster unit}
\newacronym{bw}{BW}{bandwidth}
\newacronym{ca}{CA}{carrier emitter}
\newacronym{cad}{CAD}{channel activity detection}
\newacronym{cars}{CARS}{calibration reference signal}
\newacronym{cbm}{CBM}{condition based maintenance}
\newacronym{cc}{CC}{constant current}
\newacronym{ccdf}{CCDF}{complementary cumulative distribution function}
\newacronym{ccnn}{CCNN}{circular convolutional neural network}
\newacronym{ccs}{CCS}{correlative channel sounder}
\newacronym{cdf}{CDF}{cumulative distribution function}
\newacronym{cdma}{CDMA}{code division-multiple access}
\newacronym{cdrx}{CDRX}{connected mode DRX}
\newacronym{ce}{CE}{coverage enhancement}
\newacronym{ced}{CED}{cumulative energy density}
\newacronym{ceofdm}{CE-OFDM}{constant-envelope OFDM}
\newacronym{cf}{CF}{cell-free}
\newacronym{cfmmimo}{CF-mMIMO}{cell-free massive MIMO}
\newacronym{cfo}{CFO}{carrier frequency offset}
\newacronym{cir}{CIR}{channel impulse response}
\newacronym{cla}{CLA}{closed-loop approach}
\newacronym{clk}{CLK}{clock}
\newacronym{cmos}{CMOS}{complementary metal oxide semiconductor}
\newacronym{cnn}{CNN}{convolutional neural network}
\newacronym{co}{CO}{combinatorial optimization}
\newacronym{cordic}{CORDIC}{coordinate rotation digital computer}
\newacronym{cost}{COST}{commercial off-the-shelf}
\newacronym{cots}{COTS}{commercial off-the-shelf}
\newacronym{cp}{CP}{cyclic prefix}
\newacronym{cpe}{CPE}{common phase error}
\newacronym{cpfsk}{CPFSK}{continuous phase frequency shift keying}
\newacronym{cpm}{CPM}{continuous phase modulation}
\newacronym{cpt}{CPT}{capacitive power transfer}
\newacronym{cpu}{CPU}{central-processing unit}
\newacronym{cpw}{CPW}{coplanar waveguide}
\newacronym{cqi}{CQI}{channel quality indicator}
\newacronym{cr}{CR}{coding rate}
\newacronym{crc}{CRC}{cyclic redundancy check}
\newacronym{crlb}{CRLB}{Cram\'er-Rao lower bound}
\newacronym{crs}{CRS}{cell reference signal}
\newacronym{cs}{CS}{compressed sensing}
\newacronym{cse}{CSE}{channel state estimation}
\newacronym{csi}{CSI}{channel state information}
\newacronym{csp}{CSP}{contact service point}
\newacronym{css}{CSS}{chirp spread spectrum}
\newacronym{cu}{CU}{central unit}
\newacronym{cv}{CV}{constant voltage}
\newacronym{cw}{CW}{continuous wave}
\newacronym{d2d}{D2D}{device-to-device}
\newacronym{dab}{DAB}{Digital Audio Broadcasting}
\newacronym{dac}{DAC}{digital-to-analog converter}
\newacronym{daq}{DAQ}{data acquisition system}
\newacronym{das}{DAS}{distributed antenna systems}
\newacronym{dbpsk}{DBPSK}{Differential Binary Phase Shift Keying}
\newacronym{dc}{DC}{direct current}
\newacronym{dcc}{DCC}{dynamic cooperation clustering}
\newacronym{ddc}{DDC}{digital down conversion}
\newacronym{de}{DE}{drain efficiency}
\newacronym{dft}{DFT}{discrete Fourier transform}
\newacronym{dftsofdm}{DFT-s-OFDM}{discrete Fourier Transform spread OFDM}
\newacronym{dftsofdmfdss}{DFT-s-OFDM-FDSS}{DFT-s-OFDM with frequency-domain spectral shaping}
\newacronym{dftsofdmfdssse}{DFT-s-OFDM-FDSS-SE}{DFT-s-OFDM with FDSS and spectral extension}
\newacronym{dl}{DL}{downlink}
\newacronym{dlc}{DLC}{Distributed Laser Charging}
\newacronym{dli}{DLI}{direct link interference}
\newacronym{dlt}{DLT}{Distributed Ledger Technology}
\newacronym{dm}{DM}{diffuse multipath}
\newacronym{dma}{DMA}{Direct Memory Access}
\newacronym{dmac}{DMAC}{Direct Memory Access Controller}
\newacronym{dmc}{DMC}{diffuse multipath component}
\newacronym{dmimo}{D-MIMO}{distributed MIMO}
\newacronym{dnn}{DNN}{deep neural network}
\newacronym{doa}{DOA}{direction-of-arrival}
\newacronym{dp}{DP}{differencial privacy}
\newacronym{dpd}{DPD}{digital pre-distortion}
\newacronym{dpdk}{DPDK}{Data Plane Development Kit}
\newacronym{dram}{DRAM}{dynamic random-access memory}
\newacronym{drcs}{$\Delta$RCS}{differential-radar cross section}
\newacronym{drx}{DRX}{Discontinuous Reception Mode}
\newacronym{dsb}{DSB}{double-sideband}
\newacronym{dsl}{DSL}{digital subscriber line}
\newacronym{dsp}{DSP}{digital signal processing}
\newacronym{dss}{DSS}{Dataset Storage Standard}
\newacronym{duc}{DUC}{digital up-converter}
\newacronym{dvb}{DVB}{Digital Video Broadcasting}
\newacronym{e2e}{E2E}{end-to-end}
\newacronym{easa}{EASA}{European Union Aviation Safety Agency}
\newacronym{ebg}{EBG}{electromagnetic bandgap}
\newacronym{ebw}{EBW}{excess bandwidth}
\newacronym{ec}{EC}{European Commission}
\newacronym{ecc}{ECC}{elliptic curve cryptography}
\newacronym{ecdf}{eCDF}{empirical cumulative distribution function}
\newacronym{ecdlp}{ECDLP}{elliptic curve discrete logarithm problem}
\newacronym{ecsp}{ECSP}{edge computing service point}
\newacronym{edlc}{EDLC}{electrostatic double-layer capacitors}
\newacronym{edrx}{eDRX}{Extended Discontinuous Reception Mode}
\newacronym{ee}{EE}{energy efficiency}
\newacronym{egprs}{EGPRS}{Enhanced Data Rates for GSM Evolution}
\newacronym{eh}{EH}{energy harvesting}
\newacronym{eirp}{EIRP}{equivalent isotropically radiated power}
\newacronym{em}{EM}{electromagnetic}
\newacronym{embb}{eMBB}{enhanced Mobile Broadband}
\newacronym{en}{EN}{energy neutral}
\newacronym{end}{END}{energy-neutral device}
\newacronym{enob}{ENOB}{effective number of bits}
\newacronym{eol}{EoL}{end of life}
\newacronym{ep}{EP}{energy profiler}
\newacronym{epd}{EPD}{electronic paper display}
\newacronym{epu}{EPU}{edge processing unit}
\newacronym{er}{ER}{Energy Receiver}
\newacronym{erp}{ERP}{effective radiation power}
\newacronym[plural=ESCs,firstplural=electronic speed controllers (ESCs)]{esc}{ESC}{electronic speed control}
\newacronym{esd}{ESD}{electrostatic discharge}
\newacronym{esl}{ESL}{electronic shelf label}
\newacronym{esr}{ESR}{equivalent series resistance}
\newacronym{et}{ET}{Energy Transmitter}
\newacronym{etsi}{ETSI}{European Telecommunications Standards Institute}
\newacronym{evd}{EVD}{eigenvalue decomposition}
\newacronym{evm}{EVM}{Error Vector Magnitude}
\newacronym{ewlb}{eWLB}{Embedded Wafer Level Ball Grid Array}
\newacronym{fa}{FA}{federation anchor}
\newacronym{fair}{FAIR}{Findability, Accessibility, Interoperability, and Reuse of digital assets}
\newacronym{fcf}{FCF}{frequency correlation function}
\newacronym{fd}{FD}{front-haul distance}
\newacronym{fdd}{FDD}{frequency-division duplexing}
\newacronym{fde}{FDE}{frequency domain equalizer}
\newacronym{fdm}{FDM}{frequency-division multiplexing}
\newacronym{fdma}{FDMA}{frequency division multiple access}
\newacronym{fdss}{FDSS}{frequency-domain spectral shaping}
\newacronym{fem}{FEM}{finite element analysis}
\newacronym{fembb}{feMBB}{further enhanced mobile broadband}
\newacronym{fft}{FFT}{fast Fourier transform}
\newacronym{fh}{FH}{fronthaul}
\newacronym{fhss}{FHSS}{frequency hopping spread spectrum}
\newacronym{fifo}{FIFO}{First In, First Out}
\newacronym{fim}{FIM}{Fisher information matrix}
\newacronym{fits}{FITS}{Flexible Image Transport System}
\newacronym{fl}{FL}{federated learning}
\newacronym{fom}{FoM}{figure of merit}
\newacronym{fov}{FOV}{field of view}
\newacronym{fpga}{FPGA}{field-programmable gate array}
\newacronym{fr1}{FR1}{frequency range 1}
\newacronym{fr2}{FR2}{frequency range 2}
\newacronym{fram}{FRAM}{Ferroelectric Random Access Memory}
\newacronym{fsk}{FSK}{frequency shift keying}
\newacronym{fspl}{FSPL}{free space path loss}
\newacronym{fss}{FSS}{frequency selective surface}
\newacronym{gb}{GB}{grant-based}
\newacronym{gdpr}{GDPR}{general data protection regulation}
\newacronym{gf}{GF}{grant-free}
\newacronym{gmsk}{GMSK}{Gaussian minimum-shift keying}
\newacronym{gnb}{gNB}{Next Generation Node B}
\newacronym{gni}{GNI}{gross national income}
\newacronym{gnn}{GNN}{graph neural network}
\newacronym{gnss}{GNSS}{global navigation satellite system}
\newacronym{gpclk}{GPCLK}{general purpose clock}
\newacronym{gpio}{GPIO}{General-Purpose Input/Output}
\newacronym{gpl}{GPL}{GNU General Public License}
\newacronym{gprs}{GPRS}{General Packet Radio Services}
\newacronym{gps}{GPS}{Global Positioning System}
\newacronym{gpu}{GPU}{graphical processing unit}
\newacronym{grc}{GRC}{GNU Radio Companion}
\newacronym{gscm}{GSCM}{geometry‐based stochastic model}
\newacronym{gsm}{GSM}{Global System for Mobile Communications}  %
\newacronym{gspm}{GSpM}{Generalized spatial modulation}
\newacronym{gwp}{GWP}{Global Warming Potential}
\newacronym{harq}{HARQ}{hybrid automatic repeat request}
\newacronym{hat}{HAT}{hardware attached on top}
\newacronym{hcs}{HCS}{human-centric services}
\newacronym{hdf5}{HDF5}{Hierarchical Data Format version 5}
\newacronym{hfss}{HFSS}{High Frequency Simulator Software}
\newacronym{hmd}{HMD}{head-mounted display}
\newacronym{hpbm}{HPBM}{half power beam width}
\newacronym{HyMPRo}{HyMPRo}{Hybrid Multi-Path Routing algorithm}
\newacronym{i.i.d.}{i.i.d.}{independent and identically distributed}
\newacronym{i2c}{I2C}{Inter-Integrated Circuit}
\newacronym{iaq}{IAQ}{Indoor Air Quality}
\newacronym{ib}{IB}{in-band}
\newacronym{ibbc}{IBBC}{inter-band beam configuration}
\newacronym{ibo}{IBO}{input back-off}
\newacronym{ic}{IC}{integrated circuit}
\newacronym{iccs}{ICCS}{Ilmsens correlative channel sounder}
\newacronym{ici}{ICI}{intercarrier interference}
\newacronym{icnirp}{ICNIRP}{International Commission on Non-Ionizing Radiation Protection}
\newacronym{id}{ID}{information decoding}
\newacronym{idft}{IDFT}{inverse discrete Fourier transform}
\newacronym{idxm}{IDXM}{index modulation}
\newacronym{if}{IF}{intermediate-frequency}
\newacronym{ifft}{IFFT}{inverse fast-Fourier-transform}
\newacronym{iid}{i.i.d.}{independently and identically distributed}
\newacronym{iis}{IIS}{integrated information system}
\newacronym{im}{IM}{intermodulation}
\newacronym{imd}{IMD}{intermodulation distortion}
\newacronym{imu}{IMU}{inertial measurement unit}
\newacronym{inh}{InH}{indoor hotspot office}
\newacronym{io}{IO}{input/output}
\newacronym{ioe}{IoE}{Internet of Everything}
\newacronym{iot}{IoT}{Internet of Things}
\newacronym{ipt}{IPT}{inductive power transfer}
\newacronym{ipy}{IPY}{Interventions per Year}
\newacronym{iq}{IQ}{in-phase and quadrature}
\newacronym{iqi}{IQI}{IQ imbalance}
\newacronym{ir}{IR}{infrared}
\newacronym{isi}{ISI}{intersymbol interference}
\newacronym{ism}{ISM}{industrial, scientific and medical}
\newacronym{isp}{ISP}{internet service provider}
\newacronym{itu}{ITU}{International Telecommunication Union}
\newacronym{jesd}{JESD}{Joint Electron Devices Engineering Council}
\newacronym{jfet}{JFET}{junction field effect transistor}
\newacronym{kpi}{KPI}{key performance indicator}
\newacronym{ktofdm}{KT-DFT-s-OFDM}{known-tail-DFT-s-OFDM}
\newacronym{kvi}{KVI}{key value indicator}
\newacronym{larva}{LARVA}{LARge Virtual Array}
\newacronym{lca}{LCA}{life cycle assessment}
\newacronym{lco}{LCO}{lithium cobalt oxide}
\newacronym{ldo}{LDO}{Low-dropout voltage regulator}
\newacronym{ldpc}{LDPC}{low-density parity-check}
\newacronym{led}{LED}{Light Emitting Diode}
\newacronym{less}{LESS}{Low Energy Scheduler Solution}
\newacronym{lfp}{LFP}{lithium iron phosphate}
\newacronym{lib}{LIB}{Lithium-Ion Battery}
\newacronym{lic}{LIC}{lithium-ion capacitor}
\newacronym{lid}{LID}{Lithium Iron Disulfide}
\newacronym{lidar}{LiDAR}{light detection and ranging}
\newacronym{liion}{Li-ion}{lithium-ion}
\newacronym{lipo}{LiPo}{lithium polymer}
\newacronym{lis}{LIS}{large intelligent surface}
\newacronym{llh}{LLH}{log-likelihood}
\newacronym{lls}{LLS}{link-level simulation}
\newacronym{lmd}{LMD}{Lithium Manganese Dioxide}
\newacronym{lmmse}{LMMSE}{least minimum mean square error}
\newacronym{lmo}{LMO}{lithium ion manganese oxide}
\newacronym{lna}{LNA}{low-noise amplifier}
\newacronym{lo}{LO}{local oscillator}
\newacronym{lora}{LoRa}{long range}
\newacronym{lorawan}{LoRaWAN}{long-range wide-area network}
\newacronym{los}{LoS}{line-of-sight}
\newacronym{lp}{LP}{linear programming}
\newacronym{lpf}{LPF}{low-pass filter}
\newacronym{lpt}{LPT}{laser power transfer}
\newacronym{lpwa}{LPWA}{Low Power Wide Area}
\newacronym{lpwan}{LPWAN}{low-power wide-area network}
\newacronym{lpwans}{LPWANs}{Low-Power Wide-Area Networks}
\newacronym{lqi}{LQI}{link quality indicator}
\newacronym{lrelu}{LReLU}{leaky rectified linear unit}
\newacronym{lrt}{LRT}{likelihood-ratio test}
\newacronym{ls}{LS}{least squares}
\newacronym{lsa}{LSA}{large synthetic array}
\newacronym{lsf}{LSF}{large-scale fading}
\newacronym{lsfc}{LSFC}{large-scale fading component}
\newacronym{lstm}{LSTM}{Long Short-Term Memory}
\newacronym{ltc}{LTC}{lithium thionyl chloride}
\newacronym{lte}{LTE}{Long Term Evolution}
\newacronym{lti}{LTI}{linear time-invariant}
\newacronym{lto}{LTO}{lithium titanate}
\newacronym{lusta}{LUSTA 5G }{Logistique mUltimodale Sécuritaire Téléopérée \& Autonome 5G}
\newacronym{m2m}{M2M}{machine to machine}
\newacronym{mac}{MAC}{Medium Access Control}
\newacronym{mate}{MATE}{millimeter-wave MIMO testbed}
\newacronym{mc}{MC}{Monte Carlo}
\newacronym{mcl}{MCL}{Maximum Coupling Loss}
\newacronym{mcs}{MCS}{modulation and coding scheme}
\newacronym{mcu}{MCU}{microcontroller unit}
\newacronym{mec}{MEC}{multi-access edge computing}
\newacronym{mems}{MEMS}{micro-electromechanical systems}
\newacronym{mf}{MF}{matched filter}
\newacronym{mimo}{MIMO}{multiple-input multiple-output}
\newacronym{miso}{MISO}{multiple-input single-output}
\newacronym{ml}{ML}{machine learning}
\newacronym{mlp}{MLP}{multilayer perceptron}
\newacronym{mmic}{MMIC}{monolithic microwave integrated circuit}
\newacronym{mmimo}{mMIMO}{massive MIMO}
\newacronym{mmse}{MMSE}{minimum mean square error}
\newacronym{mmtc}{mMTC}{massive machine-typed communication}
\newacronym{mmwave}{mmWave}{millimeter wave}
\newacronym{mn}{MN}{matching network}
\newacronym{mosfet}{MOSFET}{metal-oxide semiconductor field effect transistor}
\newacronym{mpc}{MPC}{multipath component}
\newacronym{mppt}{MPPT}{maximum power point tracking}
\newacronym{mr}{MR}{maximum ratio}
\newacronym{mrc}{MRC}{maximum ratio combining}
\newacronym{mrc_em}{MRC}{maximum ratio combining}
\newacronym{mrc_EM}{MRC}{Magnetic Resonance Coupling}
\newacronym{mrt}{MRT}{maximum ratio transmission}
\newacronym{mse}{MSE}{mean square error}
\newacronym{msk}{MSK}{Minimum-Shift Keying}
\newacronym{mtc}{MTC}{Machine-Type Communication}
\newacronym{multi-rat}{Multi-RAT}{multiple radio access technology}
\newacronym{multirat}{Multi-RAT}{Multiple Radio Access Technology}
\newacronym{music}{MUSIC}{MUltiple SIgnal Classification}
\newacronym{navauwall}{NAVAUWALL}{AUtomated NAVigation in WALLonia}
\newacronym{nb}{NB}{narrowband}
\newacronym{nbiot}{NB-IoT}{narrowband IoT}
\newacronym{nca}{NCA}{nickel cobalt aluminum}
\newacronym{netcdf}{NetCDF}{Network Common Data Form}
\newacronym{nf}{NF}{noise figure}
\newacronym{nfc}{NFC}{near-field communication}
\newacronym{nfv}{NFV}{network function virtualization}
\newacronym{ngmn}{NGMN}{Next Generation Mobile Networks }
\newacronym{ni}{NI}{National Instruments}
\newacronym{nicd}{NiCd}{nikkel cadmium}
\newacronym{nimh}{NiMH}{nikkel metal hydride}
\newacronym{nlos}{NLoS}{non-line-of-sight}
\newacronym{nmc}{NMC}{nickel manganese cobalt}
\newacronym{nmos}{nMOS}{n-channel metal-oxide semiconductor}
\newacronym{nn}{NN}{neural network}
\newacronym{nnls}{NNLS}{non-negative least squares}
\newacronym{noma}{NOMA}{non-orthogonal multiple access}
\newacronym{np}{NP}{Neyman-Pearson}
\newacronym{npbch}{NPBCH}{Narrowband Physical Broadcast Channel}
\newacronym{npss}{NPSS}{Narrow Band Primay Synchronization Signal}
\newacronym{nr}{NR}{New Radio}
\newacronym{nrs}{NRS}{Narrow Band Reference Signal}
\newacronym{nsss}{NSSS}{Narrowband Secondary Synchronization Signal}
\newacronym{ntp}{NTP}{network time protocol}
\newacronym{nbw}{NBW}{noise bandwidth}
\newacronym{oai}{OAI}{OpenAirInterface} %
\newacronym{obw}{OBW}{occupied bandwidth}
\newacronym{ofdm}{OFDM}{orthogonal frequency-division multiplexing}
\newacronym{ofdma}{OFDMA}{orthogonal frequency-division multiple access}
\newacronym{ofdmim}{OFDM-IM}{OFDM with index modulation}
\newacronym{olos}{OLoS}{obstructed-line-of-sight}
\newacronym{oma}{OMA}{orthogonal multiple access}
\newacronym{oob}{OOB}{out-of-band}
\newacronym{ook}{OOK}{on-off keying}
\newacronym{opbo}{OPBO}{output power backoff}
\newacronym{oran}{O-RAN}{open radio-access network}
\newacronym{os}{OS}{operating system}
\newacronym{ota}{OTA}{over-the-air}
\newacronym{otaa}{OTAA}{over-the-air authentication}
\newacronym{p1}{P1}{Phase 1}
\newacronym{p2}{P2}{Phase 2}
\newacronym{p2p}{P2P}{point-to-point}
\newacronym{pa}{PA}{power amplifier}
\newacronym{pae}{PAE}{power-added efficiency}
\newacronym{pam}{PAM}{pulse amplitude modulation}
\newacronym{pana}{PanA}{Panel A}
\newacronym{panb}{PanB}{Panel B}
\newacronym{papr}{PAPR}{peak-to-average power ratio}
\newacronym{pc}{PC}{pilot count}
\newacronym{pcb}{PCB}{printed circuit board}
\newacronym{pcg}{PCG}{power consumption gain}
\newacronym{pcie}{PCIe}{Peripheral Component Interconnect Express}
\newacronym{pcsi}{PCSI}{perfect channel state information}
\newacronym{pd}{PD}{powered device}
\newacronym{pdcch}{PDCCH}{physical downlink control channel}
\newacronym{pdf}{PDF}{probability density function}
\newacronym{pdp}{PDP}{power delay profile}
\newacronym{pdsch}{PDSCH}{physical downlink shared channel}
\newacronym{pe}{PE}{processing element}
\newacronym{peb}{PEB}{positioning error bound}
\newacronym{per}{PER}{packet error rate}
\newacronym{pet}{PET}{privacy enhancing technology}
\newacronym{pg}{PG}{path gain}
\newacronym{pgd}{PGD}{proximal gradient descent}
\newacronym{phy}{PHY}{physical}
\newacronym{pki}{PKI}{public key infrastucture}
\newacronym{pl}{PL}{path loss}
\newacronym{pla}{PLA}{physically large array}
\newacronym{pll}{PLL}{phase-locked loop}
\newacronym[plural=PMs,firstplural=person months (PMs)]{pm}{PM}{person month}
\newacronym{pmf}{PMF}{polymer microwave fiber}
\newacronym{pmu}{PMU}{power management unit}
\newacronym{pn}{PN}{pseudo-noise}
\newacronym{po}{PO}{phase offset}
\newacronym{poc}{PoC}{proof of concept}
\newacronym{poe}{PoE}{power-over-Ethernet}
\newacronym{pps}{1PPS}{pulse per second}
\newacronym{pr}{PR}{phase reversal}
\newacronym{prb}{PRB}{Physical Resource Block}
\newacronym{prbs}{PRBs}{Physical Resource Blocks}
\newacronym{prs}{PRS}{Peripheral Reflex System}
\newacronym{ps}{PS}{Processing System}
\newacronym{psd}{PSD}{power spectral density}
\newacronym{pse}{PSE}{power sourcing equipment}
\newacronym{psk}{PSK}{phase shift keying}
\newacronym{psm}{PSM}{power saving mode}
\newacronym{pss}{PSS}{primary synchronisation signal}
\newacronym{ptp}{PTP}{precision-time protocol}
\newacronym{ptrs}{PTRS}{Phase-Tracking Reference Signals}
\newacronym{ptw}{PTW}{paging time window}
\newacronym{pv}{PV}{photovoltaic}
\newacronym{pw}{PW}{planar wavefront}
\newacronym{pwm}{PWM}{pulse width modulation}
\newacronym{qam}{QAM}{quadrature amplitude modulation}
\newacronym{qos}{QoS}{quality-of-service}
\newacronym{qpsk}{QPSK}{quadrature phase-shift keying}
\newacronym{qrrls}{QR-RLS}{QR decomposition based recursive least squares}
\newacronym{quadriga}{QuaDRiGa}{QUAsi Deterministic RadIo channel GenerAtor}
\newacronym{ra}{RA}{Random Access}
\newacronym{ram}{RAM}{random-access memory}
\newacronym{ran}{RAN}{radio access network}
\newacronym{rar}{RAR}{Random Access Response}
\newacronym{rat}{RAT}{radio access technology}
\newacronym{rb}{Rb}{Rubidium}
\newacronym{rbs}{RBS}{radio base station}
\newacronym{rbw}{RBW}{resolution bandwidth}
\newacronym{rc}{RC}{raised-cosine}
\newacronym{rcs}{RCS}{radar cross section}
\newacronym{rdl}{RDL}{redistribution layer}
\newacronym{re}{RE}{radio element}
\newacronym{relu}{ReLU}{rectified linear unit}
\newacronym{rf}{RF}{radio frequency}
\newacronym{rfeh}{RFEH}{radio frequency energy harvesting}
\newacronym{rfic}{RFIC}{radio-frequency integrated circuit}
\newacronym{rfid}{RFID}{radio frequency identification}
\newacronym{rfpt}{RFPT}{radio frequency power transfer}
\newacronym{rfsoc}{RFSoC}{Radio Frequency System-on-Chip}
\newacronym{rir}{RIR}{room impulse response}
\newacronym{ris}{RIS}{reflective intelligent surface}%
\newacronym{rllmtc}{RLLMTC}{reliable low latency machine type communication}
\newacronym{rls}{RLS}{recursive least squares}
\newacronym{rms}{RMS}{root-mean-square}
\newacronym{rmse}{RMSE}{root-mean-square error}
\newacronym{rmt}{RMT}{random matrix theory}
\newacronym{rof}{RoF}{radio-over-fiber}
\newacronym{ros}{ROS}{robot operating system}
\newacronym{rpi}{RPi}{Raspberry Pi}
\newacronym{rrc}{RRC}{Radio Resource Connection}
\newacronym{rreq}{RREQ}{route request packet}
\newacronym{rsrp}{RSRP}{Reference Signals Received Power}
\newacronym{rsrq}{RSRQ}{Reference Signal Received Quality}
\newacronym{rss}{RSS}{received signal strength}
\newacronym{rssi}{RSSI}{received signal strength indicator}
\newacronym{rtc}{RTC}{real time clock}
\newacronym{rtf}{RTF}{reader talks first}
\newacronym{rtk}{RTK}{real time kinematics}
\newacronym{rts}{RTS}{ray tracing simulator}
\newacronym{ru}{RU}{Radio Unit}
\newacronym{rv}{RV}{random variable}
\newacronym{rw}{RW}{RadioWeaves}
\newacronym{rx}{RX}{receiver}
\newacronym{rzf}{RZF}{regularized zero forcing}
\newacronym{s-parameter}{S-parameter}{scattering parameter}
\newacronym{sa}{SA}{synchronization anchor}
\newacronym{sa5}{SA}{Stand Alone}
\newacronym{sar}{SAR}{specific absorption rate}
\newacronym{sbl}{SBL}{sparse Bayesian learning}
\newacronym{sc}{SC}{single carrier}
\newacronym{scfde}{SC-FDE}{single-carrier modulation with frequency-domain-equalization}
\newacronym{scfdma}{SCFDMA}{single-carrier frequency division multiple access}
\newacronym{scs}{SCS}{sub-carrier spacing}
\newacronym{sdg}{SDG}{Sustainable Development Goal}
\newacronym{sdm}{SDM}{sigma-delta modulator}
\newacronym{sdma}{SDMA}{spatial-division multiple access}
\newacronym{sdn}{SDN}{software-defined network}
\newacronym{sdof}{SDoF}{sigma-delta over fiber}
\newacronym{sdr}{SDR}{software-defined radio}
\newacronym{se}{SE}{spectral efficiency}
\newacronym{sei}{SEI}{specific emitter identification}
\newacronym{ser}{SER}{symbol-error rate}
\newacronym{sf}{SF}{spreading factor}
\newacronym{sfn}{SFN}{single frequency network}
\newacronym{sfp}{SFP}{small form-factor pluggable}
\newacronym{sha}{SHA}{Secure Hash Algorithm}
\newacronym{SigMF}{SigMF}{Signal Metadata Format}
\newacronym{simo}{SIMO}{single-input multiple-output}
\newacronym{sinr}{SINR}{signal-to-interference-plus-noise ratio}
\newacronym{siso}{SISO}{single-input single-output}
\newacronym{slam}{SLAM}{simultaneous localization and mapping}
\newacronym{slc}{SLC}{spatial leakage suppression}
\newacronym{slerp}{SLERP}{spherical linear interpolation}
\newacronym{sma}{SMA}{SubMiniature version A}
\newacronym{smc}{SMC}{specular multipath component}
\newacronym{smps}{SMPS}{switched mode power supply}
\newacronym{sndr}{SNDR}{signal-to-noise-and-distortion ratio}
\newacronym{snidr}{SNIDR}{signal-to-noise-and-interference-and-distortion ratio}
\newacronym{snir}{SNIR}{signal-to-interference-plus-noise ratio}
\newacronym{snr}{SNR}{signal-to-noise ratio}
\newacronym{soc}{SoC}{state of charge}
\newacronym{SoC}{SoC}{System on Chip}
\newacronym{sp}{SP}{service point}
\newacronym{spdt}{SPDT}{single pole double throw}
\newacronym{spi}{SPI}{Serial Peripheral Interface}
\newacronym{spst}{SPST}{single pole single throw}
\newacronym{sram}{SRAM}{static random-access memory}
\newacronym{srd}{SRD}{short-range device}
\newacronym{srls}{SRLS}{standard recursive least squares}
\newacronym{srs}{SRS}{Sounding Reference Signal}
\newacronym{ssb}{SSB}{synchronisation signal block}
\newacronym{ssd}{SSD}{solid state drive}
\newacronym{ssq}{SSQ}{simulator sickness questionnaire}
\newacronym{steam}{STEAM}{science, technology, engineering, the arts, and mathematics}
\newacronym{svd}{SVD}{singular value decomposition}
\newacronym{sw}{SW}{spherical wavefront}
\newacronym{swipt}{SWIPT}{simultaneous wireless information and power transfer}
\newacronym{synce}{SyncE}{Synchronous Ethernet}
\newacronym{ta}{TA}{timing advance}
\newacronym{tau}{TAU}{tracking area update}
\newacronym{tcer}{TCER}{transported to consumed energy ratio}
\newacronym{tcp}{TCP}{Transmission Control Protocol}
\newacronym{tcxo}{TCXO}{temperature compensated crystal oscillator}
\newacronym{tdce}{TD-CE}{time-domain compression and expansion}
\newacronym{tdd}{TDD}{time division duplexing}
\newacronym{tdma}{TDMA}{time division-multiple access}
\newacronym{tdoa}{TDOA}{time-difference-of-arrival}
\newacronym{thz}{THz}{Terahertz}
\newacronym{tmr}{TMR}{tunnel magneto resistance}
\newacronym{to}{TO}{timing offset}
\newacronym{toa}{TOA}{time-of-arrival}
\newacronym{tof}{ToF}{time-of-flight}
\newacronym{tosm}{TOSM}{through-open-short-match}
\newacronym{tpms}{TPMS}{Tire-Pressure Monitoring System}
\newacronym{trl}{TRL}{technology readyness level}
\newacronym{trp}{TRP}{Transmission Reception Point}
\newacronym{tsn}{TSN}{time-sensitive networking}
\newacronym{ttf}{TTF}{tag talks first}
\newacronym{ttff}{TTFF}{Time To First Fix}
\newacronym{tti}{TTI}{transmission time interval}
\newacronym{ttm}{TTM}{time to market}
\newacronym{ttn}{TTN}{The Things Network}
\newacronym{tx}{TX}{transmitter}
\newacronym{uart}{UART}{Universal Asynchronous Receiver/Transmitter}
\newacronym{uav}{UAV}{unmanned aerial vehicle}
\newacronym{uc}{UC}{use case}
\newacronym{ucie}{UCIe}{Universal Chiplet Interconnect Express}
\newacronym{udp}{UDP}{User Datagram Protocol}
\newacronym{ue}{UE}{user equipment}
\newacronym{ugv}{UGV}{unmanned ground vehicle}
\newacronym{uhd}{UHD}{USRP hardware driver}
\newacronym{uhf}{UHF}{ultra-high frequency}
\newacronym{ul}{UL}{uplink}
\newacronym{ula}{ULA}{uniform linear array}
\newacronym{uMIMO}{$\mu$-MIMO}{ultra-massive Multiple-Input Multiple-Output}
\newacronym{ummtc}{umMTC}{ultra massive machine type communication}
\newacronym{un}{UN}{United Nations}
\newacronym{unow}{uNOW}{unified non-orthogonal waveform}
\newacronym{upa}{UPA}{uniform planar array}
\newacronym{ura}{URA}{uniform rectangular array}
\newacronym{urllc}{URLLC}{ultra-reliable low-latency communications}
\newacronym{usrp}{USRP}{universal software radio peripheral}
\newacronym{uv}{UV}{unmanned vehicle}
\newacronym{uw}{UW}{unique word}
\newacronym{uwb}{UWB}{ultrawideband}
\newacronym{uwofdm}{UW-DFT-s-OFDM}{unique word DFT-s-OFDM}
\newacronym{v2v}{V2V}{vehicle-to-vehicle}
\newacronym{vco}{VCO}{voltage-controlled oscillator}
\newacronym{vep}{VEP}{virtual edge platform}
\newacronym{vlc}{VLC}{visible light communication}
\newacronym{vlp}{VLP}{visible light positioning}
\newacronym{vna}{VNA}{vector network analyzer}
\newacronym{voc}{VOC}{Voltatile Organic Compound}
\newacronym{votable}{VOTable}{Virtual Observatory Table}
\newacronym{vr}{VR}{virtual reality}
\newacronym{vuca}{VUCA}{volatile, uncertain, complex and ambiguous}
\newacronym{wb}{WB}{wideband}
\newacronym{wban}{WBAN}{wireless body area network}
\newacronym{wd}{WD}{Wireless distance}
\newacronym{wdt}{WDT}{watchdog timer}
\newacronym{wimax}{WiMAX}{Worldwide Interoperability for Microwave Access}
\newacronym{wlan}{WLAN}{wireless LAN}
\newacronym[plural=WPs,firstplural=work packages (WPs)]{wp}{WP}{work package}
\newacronym{wpt}{WPT}{wireless power transfer}
\newacronym{wr}{WR}{White Rabbit}
\newacronym{wrsn}{WRSN}{wireless rechargeable sensor network}
\newacronym{wsn}{WSN}{wireless sensor network}
\newacronym{xets}{XETS}{cross exponentially tapered slot}
\newacronym{xlmimo}{XL-MIMO}{extremely large-scale MIMO}
\newacronym{xr}{XR}{extended reality}
\newacronym{z3ro}{Z3RO}{zero third-order distortion}
\newacronym{zf}{ZF}{zero-forcing}
\newacronym{zmcscg}{ZMCSCG}{zero mean circularly symmetric complex Gaussian}
\newacronym{zmq}{ZMQ}{ZeroMQ}
\newacronym{ztofdm}{ZT-DFT-s-OFDM}{zero-tail DFT-s-OFDM}
\newacronym{llr}{LLR}{log-likelihood ratio}

\newacronym{cea}{CEA}{Controlled Environment Agriculture}
\newacronym{vf}{VF}{Vertical Farming}
\newacronym{dsss}{DSSS}{direct-sequence spread spectrum}
\newacronym{lbt}{LBT}{listen-before-talk}
\newacronym{afa}{AFA}{adaptive frequency agility}
\newacronym{dlpf}{DLPF}{Digital Low Pass Filter}

\input{auto_names}                  %

\usepackage{orcidlink}

\def\BibTeX{{\rm B\kern-.05em{\sc i\kern-.025em b}\kern-.08em
    T\kern-.1667em\lower.7ex\hbox{E}\kern-.125emX}}

\definecolor{color1}{HTML}{1b9e77}
\definecolor{color2}{HTML}{d95f02}
\definecolor{color3}{HTML}{7570b3}
\definecolor{color4}{HTML}{e7298a}
\definecolor{color5}{HTML}{66a61e}
\definecolor{color6}{HTML}{e6ab02}
\definecolor{color7}{HTML}{a6761d}
\definecolor{color8}{HTML}{666666}

\colorlet{AMBgreen}{color1}
\colorlet{AMBdarkgreen}{color1}
\colorlet{AMBlightgreen}{color1}
\definecolor{darkgray176}{RGB}{176,176,176}

\usepackage{tikz}
\usepackage{pgfplots}   %
\pgfplotsset{compat=newest}
\usetikzlibrary{plotmarks}
\usetikzlibrary{arrows.meta}
\usetikzlibrary{arrows}
\usetikzlibrary{calc,fadings,decorations.pathreplacing}
\usetikzlibrary{fit,backgrounds}  %

\usetikzlibrary{shapes,arrows,fit,positioning}
\usetikzlibrary{shapes}
\usetikzlibrary{spy}
\usetikzlibrary{circuits}
\usetikzlibrary{arrows}

\pgfplotsset{
    every axis/.append style={
        title style={draw=none},
        label style={font=\small},
        legend style={
            fill opacity=0.8,
            nodes={scale=0.8, transform shape}, {draw=none}
        },
        tick align=outside,
        tick pos=left,
        x grid style={darkgray176},
        xtick style={color=black},
        y grid style={darkgray176},
        ytick style={color=black},
        grid=both,
    },
    every axis plot/.append style={
        line width=2.0pt,
    },
}

\tikzset{%
  >=latex,
  inner sep=0pt,%
  outer sep=2pt,%
  mark coordinate/.style={inner sep=0pt,outer sep=0pt,minimum size=3pt,
  fill=black,circle}%
}

\usepackage{pgf-pie}

\usepackage[font=footnotesize]{caption}
\usepackage{subcaption}

\defineauthors[false]{gilles, jarne, jona, liesbet, thomas, bram, sarah}
    
\begin{document}

\title{RF-Powered Batteryless Plant Movement Sensor for Precision Agriculture
\thanks{This work is Co-funded by the European Union under Grant Agreement 101191936 (Sustain-6G), and under Grant Agreement No. 101192113. (AMBIENT-6G) and the Global Seed Fund Research Fund with No.~GSFAH/24/014 (Greenhouse Resource Optimization and Wireless Technology for Horticulture) and the KU Leuven’s Industrial Research Fund (IOF) C3 Grant ID C3/22/031.}
}

\author{
\IEEEauthorblockN{%
Jona Cappelle\,\orcidlink{0000-0002-2084-9875}\IEEEauthorrefmark{1},
Jarne Van Mulders\,\orcidlink{0000-0003-4103-3290}\IEEEauthorrefmark{1},
Sarah Goossens\,\orcidlink{0000-0001-7525-1179}\IEEEauthorrefmark{1},
Thomas Reher\,\orcidlink{0000-0002-1437-1270}\IEEEauthorrefmark{2}\IEEEauthorrefmark{3},\\
Liesbet Van der Perre\,\orcidlink{0000-0002-9158-9628}\IEEEauthorrefmark{1},
Lieven De Strycker\,\orcidlink{0000-0001-8172-9650}\IEEEauthorrefmark{1},
Bram Van de Poel\,\orcidlink{0000-0001-5638-2472}\IEEEauthorrefmark{2}\IEEEauthorrefmark{3},
Gilles Callebaut\,\orcidlink{0000-0003-2413-986X}\IEEEauthorrefmark{1}
}  
\IEEEauthorblockA{\IEEEauthorrefmark{1}%
Department of Electrical Engineering, KU Leuven, Belgium}
\IEEEauthorblockA{\IEEEauthorrefmark{2}
Department of Biosystems, KU Leuven, Belgium} %
\IEEEauthorblockA{\IEEEauthorrefmark{3}%
Leuven Plant Institute (LPI), KU Leuven, Belgium}
}

\maketitle

\begin{abstract}
Precision agriculture demands non-invasive, energy-efficient, and sustainable plant monitoring solutions. In this work, we present the design and implementation of a lightweight, batteryless plant movement sensor powered solely by \gls{rf} energy. This sensor targets \gls{cea} and utilizes \glspl{imu} to monitor leaf motion, which correlates with plant physiological responses to environmental stress. By eliminating the battery, we reduce the ecological footprint, weight, and maintenance requirements, transitioning from lifetime-based to operation-based energy storage. Our design minimizes circuit complexity while enabling flexible, adaptive readout scheduling based on energy availability and sensor data. We detail the energy requirements, \gls{rf} power transfer considerations, integration constraints, and outline future directions, including multi-antenna power delivery and networked sensor synchronization.
\end{abstract}


\glsresetall

\definecolor{pieblue}{HTML}{8ec6d7}
\definecolor{piegreen1}{HTML}{c8e7a7}
\definecolor{piered}{HTML}{f5949a}
\definecolor{piegreen2}{HTML}{b9cb96}
\definecolor{pieorange}{HTML}{f7ba7a}
\definecolor{piegrey}{HTML}{9f9f9f}
\definecolor{pieblue2}{HTML}{14A9D6}

\section{Introduction}

Horticultural practices increasingly rely on precision agriculture and smart sensing technologies to optimize plant growth conditions, maximize yields, and promptly identify stress factors~\cite{agriculture13081593}. Recent advances have demonstrated that plant movements provide insightful indicators of their physiological states, responding uniquely to various environmental stresses such as drought, salinity, and temperature fluctuations~\cite{Geldhof2021}. In this context, \gls{imu}-based wireless sensor modules have been successfully developed, offering non-intrusive monitoring capabilities suitable for long-term deployment~\cite{10784661}. These lightweight, energy-efficient sensors track plant movements and rotations.%

However, conventional battery-powered solutions, despite their effectiveness, are still constrained by limited lifetime and the environmental footprint associated with battery manufacturing, maintenance, and disposal. Batteries pose a significant environmental burden, as they must be replaced after each plant cycle and contain harmful materials~\cite{lebedeva2016considerations}. Moreover, increasing battery lifetime would require a larger battery, which conflicts with the strict weight constraints.

\begin{figure}[t]
    \hfill%
    \resizebox{1.0\linewidth}{!}{\input{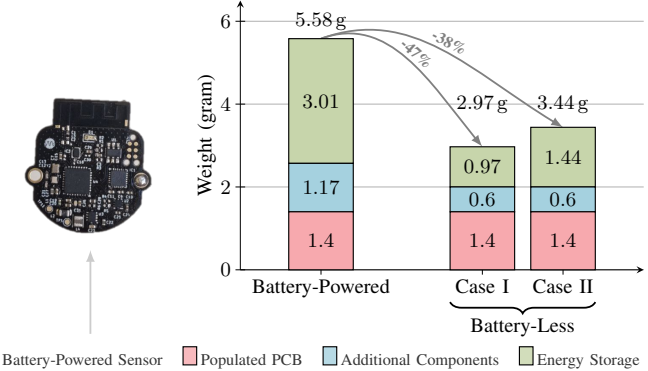}}%
    \caption{Distribution of weight according to component type for the battery-powered sensor (left) and the battery-less sensors (Case~I \& Case~II). The battery dominates the weight of the node. We can save 47\% and 38\% in the total weight by going to \acrshort{rf}-\acrshort{wpt} for Case~I, and Case~II, respectively, as elaborated in \cref{sec:rf_energy_requirement_analysis}.}
    \label{fig:weight-distribution}
\vspace{-12pt}
\end{figure}

In this paper, we propose an innovative upgrade of the wireless plant sensor (\cref{fig:weight-distribution}) to a \gls{rf}-powered, batteryless solution. This new sensor architecture eliminates reliance on traditional energy storage mechanisms, shifting from lifetime energy storage to operation-based energy storage. This significantly reduces the sensor's physical footprint and associated waste as well as its impact on plant movement interference, while theoretically indefinitely extending its operation lifetime.

The transition to a batteryless architecture offers practical and environmental advantages. Without a battery, the sensor module requires no periodic replacement or recharging, resulting in reduced manual intervention, maintenance costs, and waste. Eliminating battery management circuitry, including sleep modes and watchdog timers typically needed for battery longevity, further simplifies the sensor's electronic design. 
As the energy is now being provisioned wirelessly by the infrastructure, readouts can be made flexible and adaptive and tailored to sensor data requirements.
Measurement frequency and transmission intervals can be dynamically adjusted based on patterns detected in the processed sensor data and real-time environmental conditions, reducing or pausing power delivery during periods of low physiological activity, when fewer measurements are required, 
thus still ensuring optimal performance. %

\textbf{Contributions. }
We present a novel design for an \gls{rf}-powered batteryless plant leaf sensing solution. 
Through careful optimization, we significantly reduce the sensor's size and weight while simultaneously enhancing operations. %
We investigate sensor dimensions and weight constraints (Section~\ref{sec:application_scenario}), \gls{rf} power requirements (Section~\ref{sec:rf_energy_requirement_analysis}), and \gls{rf} power efficiency, ensuring optimal performance. Additionally, we detail the sensor design (Section~\ref{sec:design}) and discuss possible future enhancements (Section~\ref{sec:conclusion_future_work}), including multiplexing techniques, multi-antenna configurations, and synchronous sensor readouts.

\section{Application-tailored Sensing System}\label{sec:application_scenario}
The sensing platform is tailored for use in \gls{cea} environments to enable sustainable, real-time monitoring of plant health and development. Target communication and power transfer distances are in the 1--2 meter range, enabling practical deployment within typical vertical farming shelf configurations or greenhouse aisles. %
The sensor weight, including its case, is targeted to be under \SI{5}{g}, as sensors will be placed directly on leaves and excess weight can interfere with natural leaf orientation.
Dimension constraints become secondary once the form factor reaches a minimal size: straightforward manual handling dictates the acceptable limit. Cost constraints on the other hand also play an important role. \citeauthor{Moons2022} note that financial barriers (i.e., perceived high purchase and operational costs) negatively affect growers’ perceived behavioral control and thus lower adoption, with only approximately 50\% of growers %
having a high willingness to adopt new technologies~\cite{Moons2022}.

\section{RF-Energy Requirement Analysis}\label{sec:rf_energy_requirement_analysis}
A good estimate of the required energy for a leaf angle measurement is essential.
High-accuracy measurements are required as relative variation  in diurnal leaf pitch angle can be low ($\pm$ \SI{1}{\degree})~\cite{10784661}. An angular accuracy of \SI{0.1}{\degree} or better is needed to correctly analyze the leaf movement patterns, which we empirically determined and verified according to the following criteria.
Since the inherent noise of cheaper \gls{cots} \glspl{imu} is rather high, averaging is necessary to meet the criteria. For a typical example~\cite{icm20948}, the noise spectral density is \SI{230}{\micro \gram \per \sqrt{\hertz}}. Hence, the accelerometer is set up to internally average 32 samples, reducing the noise spectral density to \SI{1.91}{\milli \gram} RMS noise, and is further limited by an internal \gls{dlpf} %
with a \gls{nbw} of \SI{8.3}{Hz}. %
The compass is not capable of internal averaging, hence it is setup at the highest sampling rate (\SI{100}{Hz}) to minimize measurement time and also averaged over 32 samples. The very slow angular variations allows orientation estimation using only accelerometer and compass data, avoiding the need for a power-hungry gyroscope. 
Magnetic disturbances can easily influence the compass values, making the processed Euler angle yaw axis less accurate than the pitch and roll axes. Currently, the focus is on the pitch angle; however, future work could incorporate the roll and yaw angle, potentially leveraging machine learning models to predict plant health~\cite{agriculture13081593}.

\vspace{0.3cm}
Two scenarios are compared:
\begin{itemize}
    \item[-] \textbf{Case I:} reading only accelerometer values to determine pitch and roll leaf angles, or 
    \item[-] \textbf{Case II:} reading accelerometer and compass values, to determine all 3 Euler angles, including yaw. 
\end{itemize}

\begin{figure}[hbtp]
    \centering
    \begin{tikzpicture}

\newcommand\offset{0.003}

\pgfplotstableread[
  col sep=comma,
  ignore chars={\"}
]{figures/downsampled_RF_powered_accel_mag_32samples.csv}\datatable

\pgfplotstablegetelem{0}{[index]0}\of\datatable
\pgfmathsetmacro{\firstTimeA}{\pgfplotsretval}

\pgfplotstableread[
  col sep=comma,
  ignore chars={\"}
]{figures/downsampled_RF_powered_accel_averaged.csv}\datatableB

\pgfplotstablegetelem{0}{[index]0}\of\datatableB
\pgfmathsetmacro{\firstTimeB}{\pgfplotsretval}

  \begin{axis}[
    width=0.9\linewidth,
    height=0.6\linewidth,
    xlabel={Timestamp (ms)},
    xmin=0,
    xmax=750,
    ymin=-13,
    ylabel={Power (mW)},
    grid=major,
    legend entries={Raw (32 samples), Averaged},
    legend style={
      at={(0.5,0.98)}, %
      anchor=north,
      legend columns=3,
      draw=none,
      fill=none,
      column sep=1em,
      /tikz/every even column/.append style={column sep=0.2em}
    }
  ]

\draw[fill=red!20, draw=none] (axis cs:25,-5) rectangle (axis cs:90,0);
\draw[fill=yellow!20, draw=none] (axis cs:90,-5) rectangle (axis cs:290,0);
\draw[fill=green!20, draw=none] (axis cs:290,-5) rectangle (axis cs:320,0);

\draw[fill=red!20, draw=none] (axis cs:25,0) rectangle (axis cs:90,5);
\draw[fill=yellow!20, draw=none] (axis cs:90,0) rectangle (axis cs:695,5);
\draw[fill=green!20, draw=none] (axis cs:695,0) rectangle (axis cs:725,5);

\draw[fill=blue!20, draw=none] (axis cs:755,2) rectangle (axis cs:790,3);
\draw[fill=orange!60, draw=none]  (axis cs:755,-4) rectangle (axis cs:790,-3);

    \addplot[
      blue!20,
      thick,
      mark=*, 
      mark options={scale=0.01}
    ] table[
        x expr=(\thisrow{Timestamp}-\firstTimeA)*1000,
      y expr=\thisrow{Value}*1000,
      restrict expr to domain={\coordindex}{0:585},
      col sep=comma,
       ignore chars={\"}
    ] {./figures/downsampled_RF_powered_accel_mag_32samples.csv};
    \addlegendentry{Case II}

    \addplot[
      orange!60,
      thick,
      mark=*, 
      mark options={scale=0.01}
    ] table[
        x expr=(\thisrow{Timestamp}-\firstTimeB + \offset)*1000,
      y expr=\thisrow{Value}*1000,
      restrict expr to domain={\coordindex}{0:260},
      col sep=comma,
       ignore chars={\"}
    ] {./figures/downsampled_RF_powered_accel_averaged.csv};
    \addlegendentry{Case I}

\node[anchor=north] (initLabel) at (axis cs:100,-8) {\footnotesize Initialization};
\draw[->, thick, black!50] (initLabel.north) -- (axis cs:57.5,-2.5);

\node[anchor=north] (measLabel) at (axis cs:300,-8) {\footnotesize Measurement};
\draw[->, thick, black!50] (measLabel.north) -- (axis cs:190,-2.5);

\node[anchor=north] (wirelessLabel) at (axis cs:600,-8) {\footnotesize Wireless transmission};
\draw[->, thick, black!50] (wirelessLabel.north) -- (axis cs:305,-2.5);
\draw[->, thick, black!50] (wirelessLabel.north) -- (axis cs:710,2.5);
    
  \end{axis}
\end{tikzpicture}
    \caption{Power consumption of the leaf sensor. Stages: (1) Initialization (red), (2) Measurement (yellow), (3) Wireless transmission (green). 
    Energy per measurement: \SI{2.88}{mJ} (Case I), %
    \SI{9.11}{mJ} (Case II). %
    }
    \label{fig:power}
\end{figure}
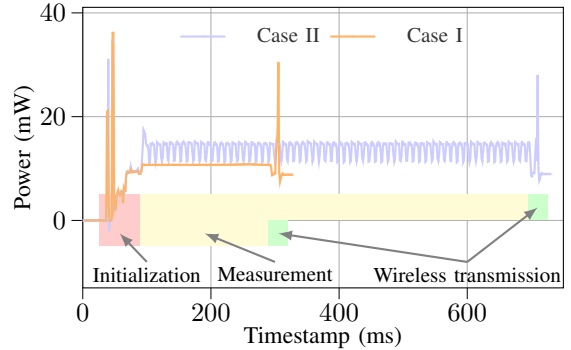

The power consumption of the leaf node, i.e., the plant sensor mounted on the leaf, is measured and depicted in \cref{fig:power} 
for one measurement and transmit cycle. A full cycle consumes \SI{2.88}{\milli\joule} of energy (Case I) or \SI{9.11}{mJ} of energy (Case II). The sequence of events is as follows: (1) Initialization of clocks, peripherals, \gls{ble} stack, powering and initialization of the \gls{imu}, (2) Measurement: requesting accelerometer and magnetometer samples, calculating the Euler angles, and (3) Wireless transmission: sending one or more \gls{ble} advertising packet(s) with the angular data. The \gls{ble} packet consists of 12 bytes at 1M PHY, sent with a transmit power of \SI{0}{dBm}.

For the \gls{rf} power transfer, we empirically determined the \gls{pl} when attaching the node to the leaves of a plant. For a transmitter at \SI{1}{\meter}, a \gls{pl} of \SI{40}{dB} can be expected. Given that a minimum of \SI{-15}{\dBm} is needed for the harvester circuit to work~\cite{harvester}, we need a transmit power of at least \SI{25}{\dBm} (\SI{-15}{\dBm} + \SI{40}{dB}).
Our solution operates in the \SI{915}{\mega\hertz} \gls{ism} band, as used by the energy harvesting chip. The maximum allowed transmit power, i.e., \gls{eirp}, is region and waveform dependent. For instance, in the United States~\cite{FCC15.247}, the maximum \gls{eirp} in the \SIrange{902}{928}{\mega\hertz} unlicensed band is \SI{36}{\dBm} (\SI{4}{\watt}). Similarly, in the European Union, in the \SIrange{915}{921}{\mega\hertz} band a maximum \gls{erp} of  \SI{36}{\dBm} is allowed \cite{EU2018_1538}. %

\section{Design}\label{sec:design}

\begin{figure}[htbp]
    \centering
    \resizebox{\linewidth}{!}{\usetikzlibrary{calc}

\begin{tikzpicture}[american voltages, every text node part/.style={align=center}, /tikz/circuitikz/tripoles/nigfetd/height=.8, /tikz/circuitikz/tripoles/nigfetd/width=.5]

\ctikzset{bipoles/resistor/height=0.25}
\ctikzset{bipoles/resistor/width=0.6}
\ctikzset{bipoles/capacitor/height=0.5}
\ctikzset{bipoles/capacitor/width=0.1}
\ctikzset{bipoles/americaninductor/height=0.35}
\ctikzset{bipoles/americaninductor/width=0.8}

\definecolor{colorSections}{rgb}{0.13, 0.55, 0.13}

\tikzset{block/.style = {rectangle, draw=black!50, fill=black!5, thick, minimum width=4cm, minimum height = 4.5cm}}
\tikzset{buck/.style = {rectangle, draw=black!50, fill=black!5, thick, minimum width=2cm, minimum height = 1.5cm}}
\tikzset{mcu/.style = {rectangle, draw=black!50, fill=black!5, thick, minimum width=3.5cm, minimum height = 4.5cm}}
\tikzset{imu/.style = {rectangle, draw=black!50, fill=black!5, thick, minimum width=2.5cm, minimum height = 3cm}}

\node [block] at (0.75,1.75) (harvester) {\Large RF harvester};
\node [buck] at (5.5,3.25) (buckconv) { \Large Buck};
\node [mcu, text depth=1.5cm] at (9.75,1.75) (stm32) {\Large BLE SoC};
\node [imu] at (15,1) (imu) {\Large IMU};

\node [] at ($ (buckconv.east) + (0.5,0.25)$) () {\large +1V9};
\node [] at ($ (buckconv.west) + (-0.75,0.25)$) () {\large +4V5};

\draw[color=black!50]
  (3.75,3.25) to[C] (3.75,-0.5) %
  (3.75, 3.25) -- (2.75, 3.25) %
  (3.75, 3.25) -- (buckconv.west) %
  (buckconv.east) -- (8, 3.25) %
;

\draw[color=black!50]
    (harvester.south) -- ++(0,-0.3)           %
    ++(0,-0.0) coordinate (a)           %
    (a) ++(-0.2,0) -- ++(0.4,0)          %
    (a) ++(-0.15,-0.1) -- ++(0.3,0)      %
    (a) ++(-0.1,-0.2) -- ++(0.2,0);      %

\draw[color=black!50]
    (buckconv.south) -- ++(0,-0.3)           %
    ++(0,-0.0) coordinate (a)           %
    (a) ++(-0.2,0) -- ++(0.4,0)          %
    (a) ++(-0.15,-0.1) -- ++(0.3,0)      %
    (a) ++(-0.1,-0.2) -- ++(0.2,0);      %

\draw[color=black!50]
    (3.75,-0.5) -- ++(0,-0.3)           %
    ++(0,-0.0) coordinate (a)           %
    (a) ++(-0.2,0) -- ++(0.4,0)          %
    (a) ++(-0.15,-0.1) -- ++(0.3,0)      %
    (a) ++(-0.1,-0.2) -- ++(0.2,0);      %
    
\draw[color=black!50]
    (stm32.south) -- ++(0,-0.3)           %
    ++(0,-0.0) coordinate (a)           %
    (a) ++(-0.2,0) -- ++(0.4,0)          %
    (a) ++(-0.15,-0.1) -- ++(0.3,0)      %
    (a) ++(-0.1,-0.2) -- ++(0.2,0);      %

\draw[color=black!50]
    (imu.south) -- ++(0,-0.3)           %
    ++(0,-0.0) coordinate (a)           %
    (a) ++(-0.2,0) -- ++(0.4,0)          %
    (a) ++(-0.15,-0.1) -- ++(0.3,0)      %
    (a) ++(-0.1,-0.2) -- ++(0.2,0);      %

\draw[draw=black!50] (-1.25,3) -- (-1.75,3)-- +(0,0.3) -- +(0.2625,0.65) -- +(-0.2625,0.65) -- +(0,0.3);

\draw[draw=black!50] (11.5,3) -- (12.0,3)-- +(0,0.3) -- +(0.2625,0.65) -- +(-0.2625,0.65) -- +(0,0.3);

\draw[color=black!50] (harvester.north) -- ($ (harvester.north) + (0,1) $) -- ($ (buckconv.north) + (0,1) $) -- (buckconv.north);

\draw[draw=black!50] (11.5,1.75) -- (13.75,1.75);

\node [] at (12.75,1.3) () {\large SPI};
\draw[draw=black!50] (11.5,0.75) -- (13.75,0.75);

\draw[draw=black!50] (12.5,0.5) -- (12.75,1);

\node [] at (12.25,4.5) () {\large 2.4 GHz\\[5pt] \large BLE \par};
\node [] at (-2,4.5) () {\large 915 MHz};

\node [] at (10.75,1.75) () {\large GPIO};

\node [] at (14.5,1.75) () {\large VDD};
\node [] at ($(imu.south) + (0,0.5)$) () {\large GND};

\node [] at ($(stm32.south) + (0,0.5)$) () {\large GND};

\node [] at (11,3) () {\large RF};

\node [] at (2.1,3.25) () {\large BATT};

\node[] at ($(harvester.south) + (0,0.5)$) {\large GND};

\node [] at (0.5,3.5) () {\large STATUS[0]};
\node [] at (-0.5,2.75) () {\large RF IN};
\node [] at (4.15,1.0) (buffer) {\large $C_{B}$};
\node [] at ($ (buckconv.north) + (0.5,0.25)$) () {\large EN};

\end{tikzpicture}%
}
    \caption{System design: An \acrshort{rf} energy harvester charges a storage capacitor $C_B$. A buck converter regulates the voltage across the capacitor to supply power to a \acrshort{ble} \acrshort{SoC}, which interfaces with an \acrshort{imu} via the \acrshort{spi} bus.}
    \label{fig:rfwpt-design}
    \vspace{-12pt}
\end{figure}
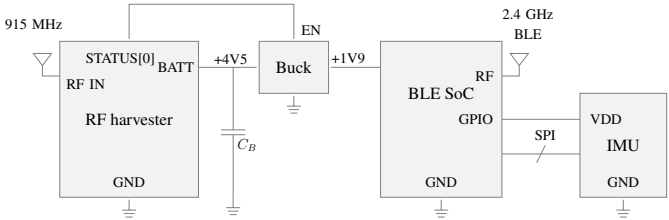

The system design is depicted in \cref{fig:rfwpt-design} and consists of the AEM30940 \gls{rf} energy harvester, a capacitor $C_B$, a buck converter, an STM32WB09 \gls{SoC} with \gls{ble} communication, and an ICM20948 \gls{imu}.
The \gls{rf} harvester gets its energy from a monopole antenna connected through a matching network and a rectifier to the \gls{ic}. The incident \gls{rf} energy charges the capacitor $C_B$. An \gls{aec} is selected for its low \gls{esr} and quick charging time. 
The buck converter ensures efficient conversion of the higher capacitor voltage (\SI{4.5}{V}) to the required voltage by the node (\SI{1.9}{V}). The energy harvester features several voltage threshold levels that can be used at the application side. The $STATUS[0]$ enable pin of the energy harvester is used to enable the buck converter when the capacitor is charged to $V_{CHRDY}$ (\SI{4.5}{V}), activating the node, or disabling the buck converter at $V_{OVDIS}$ ($\pm$\SI{2.2}{V}), deactivating the node. Alternatively, the energy harvester’s integrated \gls{ldo} could be used, however, this would significantly reduce the available energy at the node due to the inefficient conversion given the high voltage difference.
Following the design guidelines from~\cite{VanM2506:Designing}, given the minimum required capacitor voltage ($V_{OVDIS}$) to provide a stable output to the \gls{SoC}, the selected maximal charge voltage ($V_{CHRDY}$), and the required energy of \SI{2.88}{\milli\joule} or \SI{9.11}{\milli\joule} ($E_{\text{req}}$), the required energy buffer's capacity $C_{B}$ were determined as:
\begin{equation*}
  C_{\mathrm{B}}
  = \frac{2\,E_{\mathrm{req}}}{V_{\mathrm{CHRDY}}^{2} - V_{\mathrm{OVDIS}}^{2}}
  \;=\;\begin{cases}
    \SI{346}{\micro\farad}, & E_{\mathrm{req}} = \SI{2.88}{\milli\joule},\\
    \SI{1.1}{\milli\farad},  & E_{\mathrm{req}} = \SI{9.11}{\milli\joule}.
  \end{cases}
  \label{eq:buffercapacitor}
\end{equation*}
A capacity ($C_\text{B}$) of \SI{346}{\micro\farad} or \SI{1.1}{mF} is required. A capacitor with a nominal value of \SI{470}{\micro F} (measured value \SI{490}{\micro F}) for Case I and \SI{1}{mF} (measured value \SI{1.1}{mF}) for Case II have been chosen.
If, after a measurement has been completed, there is energy remaining in the buffer, i.e., the capacitor voltage is above the buck converter’s minimum operating threshold, the node continues to send \gls{ble} advertising packets to maximize the likelihood of successful data reception.

\begin{figure}[hbtp]
    \centering
    \begin{tikzpicture}

\newcommand\offset{0.003}

\pgfplotstableread[
  col sep=comma,
  ignore chars={\"}
]{figures/downsampled_RF_powered_accel_mag_32samples.csv}\datatable

\pgfplotstablegetelem{0}{[index]0}\of\datatable
\pgfmathsetmacro{\firstTimeA}{\pgfplotsretval}

\pgfplotstableread[
  col sep=comma,
  ignore chars={\"}
]{figures/downsampled_RF_powered_accel_averaged.csv}\datatableB

\pgfplotstablegetelem{0}{[index]0}\of\datatableB
\pgfmathsetmacro{\firstTimeB}{\pgfplotsretval}

  \begin{axis}[
    width=0.9\linewidth,
    height=0.45\linewidth,
xlabel={Input power [dBm]},
    legend cell align=left,
        ymin=0,
    ymax=30,
ylabel={Charge time [min]},
    grid=major,
    legend entries={Raw (32 samples), Averaged},
    legend style={
      at={(0.5,-0.35)}, %
      anchor=north,
      legend columns=2,
      draw=none,
      fill=none,
      column sep=1em,
      /tikz/every even column/.append style={column sep=0.2em}
    }
  ]

\addplot [
      orange!60,
      thick,
      mark=*, 
    mark options={scale=0.5}
]
table {%
-15 51.6674441655477
-12.5 14.2852517127991
-10 6.51599034865697
-7.5 3.03283656040827
-5 1.5777442574501
-2.5 0.794835801919301
0 0.399235741297404
};
\addlegendentry{\SI{470}{\micro F} (cold start)}

\addplot [
      blue!40,
      thick,
      mark=*, 
      mark options={scale=0.5}
]
table {%
-15 116.099621045589
-12.5 30.0775133252144
-10 13.095721968015
-7.5 6.31956293185552
-5 3.19734695752462
-2.5 1.65223197937012
0 0.86369704802831
};
\addlegendentry{\SI{1}{\milli F} (cold start)}

\addplot [
        dashed,
      orange!60,
      thick,
      mark=*, 
      mark options={scale=0.5}
]
table {%
-15 32.4160100777944
-12.5 8.42984786430995
-10 3.60942230621974
-7.5 1.63517007827759
-5 0.830230331420899
-2.5 0.426346508661906
0 0.219642905394236
};
\addlegendentry{\SI{470}{\micro F} (successive meas)}

\addplot [
        dashed,
      blue!40,
      thick,
      mark=*, 
      mark options={scale=0.5}
]
table {%
-15 77.8878621260325
-12.5 18.9325842062632
-10 7.41878706614176
-7.5 3.46779428720474
-5 1.74288170337677
-2.5 0.882898986339569
0 0.460833887259165
};
\addlegendentry{\SI{1}{\milli F} (successive meas)}
    
  \end{axis}
\end{tikzpicture}
    \caption{Measured charge time of the energy buffer $C_{B}$ under typical expected \gls{rf} input power levels. The energy harvester reaches usable cold start intervals of \SI{15}{min} at input powers of \SI{-12}{dBm} for Case~I and \SI{-10}{dBm} for Case~II.}
    \label{fig:chargeTime}
\vspace{-12pt}
\end{figure}
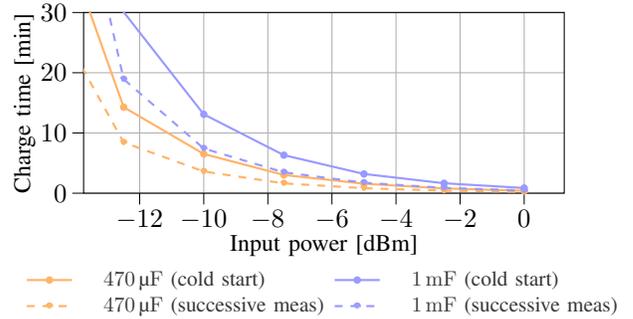

The charge time of the capacitors for different input powers has been measured.
As depicted in \cref{fig:chargeTime}, the charge time in function of input power at the energy harvester reaches usable cold start intervals of \SI{15}{min} between \SI{-12}{dBm} for Case I and \SI{-10}{dBm} for Case II. In successive measurements, the required input power reduces by \SI{1.5}{dB} for the same measurement duration (15 min), i.e., with the capacitor already charged to $V_{\mathrm{OVDIS}}$.

The current battery-powered design (\SI{5.58}{\gram}), as depicted in \cref{fig:weight-distribution}, features a \SI{1.4}{\gram} \gls{pcb} with the majority of the weight (\SI{4.18}{\gram}) contributed by the battery and the battery mount, excluding packaging.
The weight of the proposed RFPT design of the plant sensor extension can be reduced by 47\% to \SI{2.97}{g} with a capacitor of \SI{470}{\micro F} (Case~I), or to \SI{3.44}{g} with a \SI{1}{mF} capacitor (Case~II).

Comparing the sensor node to a battery-powered equivalent, the sleep mode logic and external \gls{wdt} can be removed, as the system will be intermittently powered. This reduces \gls{bom} cost and energy consumption.
Additional components needed for the \gls{rf}-powered variant are the \gls{rf} energy harvester, a rectification and matching circuit, and a dedicated antenna for harvesting. Alternatively, energy can be harvested in the \SI{2.4}{GHz} band, where an \gls{rf} switch is needed to periodically switch between energy harvesting and data transfer. Also, the maximal allowed transmit power is generally lower and more complex access techniques are required to follow the regulations in this frequency band, such as frequency hopping, listen before talk, etc.

\section{Conclusions and Future Work}\label{sec:conclusion_future_work}

We have evaluated a proposed \gls{rf}‐powered, batteryless plant movement sensor that reduces node mass by up to 47\% (\SI{2.97}{g}). The novel system could be utilized across different plant species, opening further applications in plant research, crop breeding, and large‐scale cultivation of delicate crops such as lettuce or ornamentals. The low \gls{bom} cost with \gls{cots} components 
aligns with growers’ hesitant adoption intent. By removing batteries, the node design complexity is reduced and the intelligence is shifted to the \gls{rf} infrastructure, enabling flexible readout intervals tied directly to \gls{rf} availability and sensed plant dynamics and in principle indefinitely extending its operational lifetime. %

Future work will focus on integrating novel \gls{tmr} compass or low-noise accelerometers, to improve update rates or \gls{wpt} distances.
Backscattering communication can be explored to lower the energy consumption during transmission.
To scale efficiently, strategies like frequency‐ or time‐division multiplexing schemes can be used to support simultaneous charging and communication of multiple leaf nodes, enabling spatiotemporal analysis of plant movement.
Multi-antenna power delivery in phased-array or distributed-antenna configurations can improve \gls{rf} coverage and can increase \gls{wpt} efficiency in dense canopy environments.
Measuring signal attenuation, or path-loss could enable spatial mapping of canopy density, offering a non-invasive way to track canopy development over time.

\printbibliography%

\end{document}